# Scalar Field Mapping with Adaptive High-Intensity Region Avoidance

Muzaffar Qureshi[1]    Tochukwu Elijah Ogri[1]    Zachary I. Bell[2]    Rushikesh Kamalapurkar[1]

*Abstract*— This research is motivated by a scenario where a group of UAVs is assigned to map an unknown scalar field, with the imperative of maintaining a safe distance from the sources of the field to evade detection or damage. The location of the sources is unknown a priori, so the UAVs rely on measurements of the field intensity to gauge safety. The UAVs estimate the unknown scalar field using Gaussian process (GP) regression and use the estimate to generate a map of high-intensity regions using Hough transform (HT), updated online based on the field measurements. A convergence analysis shows the boundedness of the error between the actual scalar field and the learned scalar field. The effectiveness of the method is evaluated through simulations, showcasing its ability to accurately learn scalar fields with multiple high-intensity regions while reducing the number of measurements taken inside the high-intensity regions.

## I. INTRODUCTION

Measurement and estimation of scalar fields has long been a focal point of research, as evident by a substantial body of work [1]–[4]. These research efforts focus on the creation of an accurate scalar field map that minimizes the error between the actual and the estimated scalar field with minimal time and control effort. Traditional methods for mapping scalar fields often rely on data from a sensor carried by a single agent in the field [5]–[7]. These approaches face challenges in achieving broad spatial coverage due to the limitations of the agent. Additionally, as the size of the scalar field expands, collection of comprehensive data through a single source becomes increasingly cost-prohibitive [8].

Recently, multi-agent systems have demonstrated their efficacy in learning scalar fields through distributed sensor networks, where agents operate autonomously and communicate to achieve a common objective [9]–[11]. Individual agents can also leverage measurements from neighboring agents to adjust their estimates, thereby enhancing the accuracy and the speed of scalar field estimation. Achieving consensus among agents can be challenging when they have communication constraints. Furthermore, the communication cost grows rapidly if agents are dispersed across a vast environment [12].

Due to recent advancements in computational intelligence and machine learning, learning scalar field maps with Gaussian processes (GP) has gained popularity [13]. GP methods enable us to adopt a non-parametric approach that is more robust to deviations in input data and also eliminates the need for a specific underlying function that models the scalar field. This flexibility allows GP to adapt to different field estimation problems without the necessity for extensive reconfiguration or redesign [14], [15].

In practical scenarios, the scalar fields that require mapping frequently encompass nuclear, chemical, or electromagnetic domains. The significance of these mappings extends across a variety of applications, such as environmental monitoring, detection of hostile radars, tracking fire spreads, managing nuclear outbreaks, and measuring chemical or biological concentrations [16], [17]. Mapping these fields necessitates meticulous guidance of the agents to maintain a safe distance from potential hazards.

Detection of sources in scalar fields is usually conducted using large-range sensors capable of identifying hazards from a safe distance. In [18], Aleotti et al. utilized haptic teleoperation for nuclear radiation detection. Okuyama et al. proposed a helicopter-based radiation detection strategy using radiation detectors and CCD cameras [19]. Real-world applications of these approaches face limitations since these methods necessitate frequent data transmission to a ground station, rendering them impractical for use in hostile environments with communication constraints.

Another approach is to model the hazardous regions as physical obstacles and to guide the agents to maintain a safe distance [20], [21]. This requires prior knowledge of the scalar field, which is often unavailable in unfamiliar environments.

This research addresses a research challenge that encompasses two primary objectives. The first is to generate an estimate of a scalar field; and the second is the acquisition of substantial information about potential hazards, characterized by high field intensity to keep the agents safe. Achieving this dual aim necessitates the formulation of a strategy that requires demarcation of the boundaries of high-intensity regions.

## II. PROBLEM FORMULATION

Consider $N$ agents deployed in a fixed 2-dimensional configuration space $\mathcal{X} \subset \mathbb{R}^2$, equipped with sensors to measure the intensity $\mathbf{Y} = g(\mathbf{x})$ of a scalar field $g : \mathcal{X} \to \mathbb{R}$, where $\mathbf{x}$ represents the position of the agent. The objective is to assign measurement locations, denoted by $\mathbf{x}_t^i$, to agents $i = 1, \ldots, N$ at time steps $t \in \{1, \ldots, n_{\text{trn}}\}$ to generate an estimate of the scalar field while minimizing presence in high-intensity regions of the scalar field.

This research is supported by the Air Force Research Laboratories under contract number FA8651-22-S-0006. Any opinions, findings, or recommendations in this article are those of the author(s), and do not necessarily reflect the views of the sponsoring agency.
[1] School of Mechanical and Aerospace Engineering, Oklahoma State University, email: {muzaffar.qureshi, tochukwu.ogri, rushikesh.kamalapurkar} @okstate.edu.
[2] Air Force Research Laboratories, Florida, USA, email: zachary.bell.10@us.af.mil.

Initially, the measurement locations may be assigned randomly since no prior knowledge is available. The agents go to the assigned measurement locations iteratively, the goal is to improve the estimate of the scalar field and create a high-intensity map, to adjust the subsequent measurement locations to reduce exposure to high field intensity. To avoid detection, all agents are also required to maintain their own estimates of the scalar field, with information sharing only allowed if the agents are in physical proximity with each other. This brief information exchange is necessary to effectively utilize the area coverage and also forward information about high-intensity between the agents.

The following assumptions are needed to make the scalar field learning problem tractable.

*Assumption 1:* The scalar field $g$ is differentiable and the norm of the derivative of the field is bounded by an unknown constant $L \geq 0$, such that $|\nabla g(\mathbf{x})| \leq L$, for all $\mathbf{x} \in \mathcal{X}$.

*Assumption 2:* The intensity of the field is bounded and time-invariant, i.e., $0 \leq g_{\min} \leq g(\mathbf{x}) \leq g_{\max}$, for all $\mathbf{x}$ in $\mathcal{X}$.

*Assumption 3:* The agents are assumed to be susceptible to detection or damage when operating in areas where the field intensity is higher than a user-selected threshold. In particular, a high-intensity region is defined as $\mathbf{G} := \{\mathbf{x} \in \mathcal{X} \mid g(\mathbf{x}) > g_{\text{thresh}}\}$, where $g_{\text{thresh}}$ is the user-defined threshold.

*Assumption 4:* In each time step $t$, each agent $i$ travels to its assigned measurement location $\mathbf{x}_t^i$ and measures the field intensity $y_t^i = g(\mathbf{x}_t^i)$.

*Assumption 5:* Agents can communicate their field estimates to other agents that are closer than a pre-defined communication radius $R_c$.

## III. METHODOLOGY

In this paper, a new approach for the detection and avoidance of high-intensity regions in scalar fields is developed using a combination of Gaussian process (GP) regression and the Hough transform (HT). Initially, GP regression is used to predict the scalar field values for the complete field, including the high-intensity regions. The challenge to create a high-intensity map is solved using the HT to accurately predict the size and the locations of the high-intensity areas. The high-intensity map is then used by the agents to adjust their future measurement locations to minimize high-intensity exposure. In the following sections, the developed methodology is described in detail.

### A. Gaussian Process Regression Model

GP regression is a probabilistic modeling technique that can be used for learning nonlinear functions [13]. Similar to a Gaussian distribution over a random variable, a GP distribution over functions is described by a mean and a covariance function. The objective of each agent $i$ is to generate its field model $\bar{g}^i(\mathbf{x})$ using GP regression given as

$$\bar{g}^i(\mathbf{x}) \sim \mathrm{GP}(\mu^i(\mathbf{x}), k^i(\mathbf{x}, \mathbf{x}')), \quad \forall i = 1, 2, \ldots, N, \quad (1)$$

where $\mu^i(\mathbf{x}) : \mathbb{R}^2 \to \mathbb{R}$ represents the mean function and $k^i(\mathbf{x}, \mathbf{x}') : \mathbb{R}^2 \times \mathbb{R}^2 \to \mathbb{R}$ represents the covariance function.

For this research, a square exponential covariance kernel with fixed hyperparameters is used, given by

$$k(\mathbf{x}, \mathbf{x}', \theta) = \alpha^2 \exp\left(-\frac{||\mathbf{x} - \mathbf{x}'||^2}{2\beta^2}\right). \quad (2)$$

The kernel hyperparameters are $\theta = \{\alpha, \beta\} \in (0, \infty)^2$ where $\alpha$ is the variance parameter that scales the output, and $\beta$ is the length-scale parameter that controls the correlation between the points. At time step $t$, the vector containing the field intensity values measured so far by agent $i$ at measurement locations $\mathbf{X}_{trn}^i(t) := \{\mathbf{x}_j^i\}_{j=1}^t$ is denoted by $\mathbf{Y}_{trn}^i(t) := \{y_j^i\}_{j=1}^t$, where $y_j^i = g(\mathbf{x}_j^i)$.

To gauge the accuracy of the predicted scalar field, we also define a set of uniformly distributed testing locations $\mathbf{X}_{test} := \{\mathbf{x}_1, \mathbf{x}_2, \ldots, \mathbf{x}_{n_{test}}\}$ in $\mathcal{X}$, where $n_{test} \gg n_{trn}$. The training kernel matrix $K_{trn-trn}(t) \in \mathbb{R}^{n_{trn} \times n_{trn}}$ at time step $t$, the test-train kernel matrix $K_{test-trn}(t) \in \mathbb{R}^{n_{test} \times n_{trn}}$ at time step $t$, and the constant testing kernel matrix $K_{test-test} \in \mathbb{R}^{n_{test} \times n_{test}}$ can now be defined using

$$K_{trn-trn}(t)(j,k) := k(\mathbf{x}_j, \mathbf{x}_k), \forall \mathbf{x}_j, \mathbf{x}_k \in \mathbf{X}_{trn}(t), \quad (3)$$

$$K_{test-test}(j,k) := k(\mathbf{x}_j, \mathbf{x}_k), \forall \mathbf{x}_j, \mathbf{x}_k \in \mathbf{X}_{test}, \quad (4)$$

$$K_{test-trn}(t)(j,k) := k(\mathbf{x}_j, \mathbf{x}_k),$$
$$\forall \mathbf{x}_j \in \mathbf{X}_{test} \text{ and } \forall \mathbf{x}_k \in \mathbf{X}_{trn}(t). \quad (5)$$

Since the agents need to navigate through the an unknown scalar field, an iterative GP model is developed, where at each timestep, one measurement is added to the GP model. As a result, the corresponding dimensions of $\mathbf{Y}_{trn}^i(t)$, $K_{trn-trn}(t)$, and $K_{test-trn}(t)$ grow with time.

The measurement locations are unique to each agent, however the same set of testing locations is used for all agents. The total number of measurement locations $n_{trn}$ (same as the total number of time steps) and testing locations $n_{test}$ are fixed for all agents. Given a set of testing locations $\mathbf{X}_{test}$, the mean vector $\mathbf{M}_t^i$ and the covariance matrix $\mathbf{S}_t^i$ as computed by agent $i$ at time $t$, from the GP posterior can be calculated as

$$\mathbf{M}_t^i = K_{test-trn}^i(t) \cdot (K_{trn-trn}^i(t))^{-1} \cdot \mathbf{Y}_{trn}^i(t), \quad (6)$$

and

$$\mathbf{S}_t^i = K_{test-test}^i - K_{test-trn}^i(t) \cdot (K_{trn-trn}^i(t))^{-1} \cdot (K_{test-trn}^i(t))^\top. \quad (7)$$

### B. Detection of High-Intensity Regions Using Hough Transform

In this section, a technique to create high-intensity maps concurrently with the GP model using the HT is developed. HT is a mathematical technique used in image analysis, extensively applied for the detection of geometric figures, primarily lines, circles, and other shapes, within an image [22]. The essence of HT lies in its transformation of the image space into a parameter space where geometric shapes are easier to identify. The implementation of the HT involves first generating a binary map of the predicted field using the value $g_{\text{thresh}}$, a user-defined threshold, that specifies the maximum intensity level that is detrimental to the agent. This binary map is obtained by applying a threshold filter to the GP estimates at time $t$, resulting in a matrix $\mathcal{G}_t^i \in \mathbb{R}^{n_{\text{test}} \times n_{\text{test}}}$ defined as:

$$\mathcal{G}_t^i(j,k) = \begin{cases} 0 & \text{if } \mathbf{M}_t^i(j,k) \leq g_{thresh}, \text{ for } j,k=1,...,n_{\text{test}}, \\ 1 & \text{Otherwise.} \end{cases} \quad (8)$$

Upon applying the circular HT to this binary image, the algorithm identifies circles, defined by their center coordinates and radii, that best approximate the high-intensity regions at time $t$. The process results in a set of circles:

$$\mathcal{C}_t^i := \{(x_j, y_j, r_j)_t \mid j = 1, \ldots, p_t\}, \quad (9)$$

where $(x_j, y_j)$ represents the center coordinates of the $j$-th circle, $r_j$ represents its radius, and $p_t$ represents the total number of circles detected at timestep $t$ for agent $i$.

### C. Avoidance of High-Intensity Regions

Through the integration of GP and HT algorithms, agents can estimate the location and size of high-intensity regions and readjust their measurement locations. At each time step $t$, the agents compute the mean $\mathbf{M}_t^i$ and covariance $\mathbf{S}_t^i$, along with the set $\mathcal{C}_t^i$ containing the circles that demarcate the high-intensity regions.

The unvisited future measurement locations undergo an evaluation for presence inside the circles using the coordinates and radii data derived from the GP-HT algorithm. For the measurement locations inside high-intensity circles, an adjustment is carried out, that involves the addition or subtraction of the radius of the circle from one of the coordinates of the initially planned measurement location to ensure the measurement location is re-positioned outside the periphery of the circle. Given the stochastic nature of this reassignment and the existence of multiple high-intensity circles, the new measurement location may inadvertently be placed within another high-intensity circle. To mitigate this, the relocation process is repeated until the new measurement location does not intersect with any high-intensity circle.

Measurement locations that are found to be outside of the high-intensity circles at time $t$ are not moved in that time step. However, they are subject to re-evaluation in subsequent time steps due to variation in the set $\mathcal{C}_t^i$.

### D. Information Sharing among Agents

Since the overall objective is to avoid detection, the agents are restricted to short-range communication. A decentralized methodology is adopted where each agent learns its own scalar field model, and two agents only communicate when they are close to each other. Contrary to approaches that rely on long-range connectivity among agents to enhance the accuracy and learning rate of a centralized GP model, the developed methodology prioritizes the safety of agents.

The topology of communication between the agents is defined using a graph $\mathbb{G}(N, E)$, where $N$ is the set of agents and $E$ is the set of edges connecting the agents. Each agent can communicate its mean vector $\mathbf{M}_t^i$ and the covariance matrix $\mathbf{S}_t^i$ with other agents in a communication radius $R_c$ at any given time $t$ and updates its GP model using the following averaging model

$$\mathbf{M}_t^i = \mathbf{M}_t^j = \frac{\mathbf{M}_t^i + \mathbf{M}_t^j}{2}, \; \mathbf{S}_t^i = \mathbf{S}_t^j = \frac{\mathbf{S}_t^i + \mathbf{S}_t^j}{2}, \quad (10)$$

if $i \neq j$ and $\|\mathbf{x}_i - \mathbf{x}_j\| \leq r_c$, where $i, j \in 1, 2, \ldots, N$.

The averaging model employed for data sharing between the agents can be optimized to be cognizant of the predictive covariance of the estimates generated by the individual agents. By assigning higher weights to agents with lower predictive uncertainty, the overall accuracy of the decentralized GP models could potentially be enhanced. However, development and evaluation of data-sharing strategies is beyond the scope of this paper and will be addressed in future work.

---

**Algorithm 1** GP-HT Algorithm

**Input**: Scalar function $g(\mathbf{x})$, GP parameters $\alpha$ and $\beta$, kernel function $k$, threshold value $g_t$
**Initialize**: Initial random measurement locations $(\mathbf{X}_{trn}^i)$ and testing $(\mathbf{X}_{test}^i)$ locations for all $N$ agents in $\mathcal{X}$
**for** $t = 1, 2, \ldots, n_{trn}$ **do**
  **for** $i = 1, 2, \ldots, N$ **do**
    Move agent to $\mathbf{x}_t^i$, measure $y_t^i$ and update $\mathbf{Y}_{trn}^i$
    Calculate $\mathbf{M}_t^i$ and $\mathbf{S}_t^i$ (6 & 7)
    Create binary map $\mathcal{G}^i$ (8)
    Create set $\mathcal{C}^i$ (9)
    **for** $m = 1, 2, \ldots, p$ **do**
      **for** $n = 1, 2, \ldots, (n_{trn} - t)$ **do**
        **if** $\mathbf{x}_i^{t+n}$ lies inside $C_m^i$ **then**
          Replace $\mathbf{x}_{t+n}^i$
        **end if**
      **end for**
    **end for**
  **end for**
  **for** $j = 1, 2, \ldots, N - 1$ **do**
    **if** $\|\mathbf{x}_t^i - \mathbf{x}_t^j\| \leq R_c$ **then**
      Average GP model $\mathbf{M}_t^i$, $\mathbf{S}_t^j$ (10)
    **end if**
  **end for**
  Compute error $\|g(\mathbf{x_{test}}) - \mathbf{M}_t^i\|_2$ at step $t$
  Move the agent to $\mathbf{x}_{k+1}^i$ location
**end for**
**Output**: $\{(\mathbf{M}^i, \mathbf{S}^i \text{ and } \mathcal{C}^i) \mid i = 1, \ldots, N\}$

---

## IV. CONVERGENCE ANALYSIS

This section focuses on the proof of convergence of the GP estimates $\bar{g}_i$ to $g$ given fixed hyperparameters $(\alpha, \beta)$ and varying measurement locations. The convergence analysis is required to prove the boundedness of the error between the actual scalar field and the predicted scalar field with varying $\mathbf{X}_{trn}^i$. Given the set of measurement locations $\mathbf{X}_{trn}^i \subseteq \mathcal{X}$, we follow [23] and define the fill distance $h_{X_{trn}^i, \mathcal{X}}$, separation radius $q_{X_{trn}^i, \mathcal{X}}$, and mesh ratio $\rho_{X_{trn}^i, \mathcal{X}}$ as

$$h_{\mathbf{X}_{trn}^i, \mathcal{X}} := \sup_{\mathbf{x} \in \mathcal{X}} \inf_{\mathbf{y} \in \mathbf{X}_{trn}^i} \|\mathbf{x} - \mathbf{y}\|, \quad (11)$$

$$q_{\mathbf{X}_{trn}^i, \mathcal{X}} := \frac{1}{2} \min_{j \neq k} \|\mathbf{x}_j^i - \mathbf{x}_k^i\|, \quad \forall j, k = 1, 2, \ldots n_{trn}, \quad (12)$$

$$\rho_{\mathbf{X}_{trn}^i, \mathcal{X}} := \frac{h_{\mathbf{X}_{trn}^i, \mathcal{X}}}{q_{\mathbf{X}_{trn}^i, \mathcal{X}}} \geq 1. \quad (13)$$

The fill distance is the maximum possible distance between any two points in $\mathcal{X}$ and $\mathbf{X}_{trn}^i$, and the separation radius is half the smallest distance between any two distinct points in $\mathbf{X}_{trn}^i$. The mesh ratio is the ratio of these quantities. The three quantities above provide measures of how uniformly the measurement locations $\mathbf{X}_{trn}^i$ are distributed in $\mathcal{X}$.

*Lemma 1:* [24, Corollary 10.48] Given a bounded Lipschitz continuous domain $\mathcal{X}$, and a compact set $S \subseteq (0, \infty)^2$ such that covariance parameters $\theta := \{\alpha, \beta\}$ satisfy $\theta \in S$, the norms associated with the native space and the Sobolev space are equivalent. That is, there exists constants $C_1$ and $C_2$ such that

$$C_1(\theta) \|\bar{g}\|_{H_{k_\theta}(\mathcal{X})} \leq \|\bar{g}\|_{H^{\frac{\nu+2}{2}}(\mathcal{X})} \leq C_2(\theta) \|\bar{g}\|_{H_{k_\theta}(\mathcal{X})}$$

In the lemma above, $\nu \geq 0$ controls the smoothness of the kernel. The square exponential kernel is the limiting case as $\nu$ approaches infinity. For the convergence of $\bar{g}_i$ to $g$, we invoke [23, Theorem 3.5]. For fixed given vector of hyperparameters $\theta \in \mathbb{R}^2$ if

1) $\mathcal{X} \subseteq \mathbb{R}^2$ is compact, with a Lipschitz continuous boundary,
2) the native space $H_{k_\theta}(\mathcal{X})$ is the Sobolev space $H^{\frac{\nu+2}{2}}(\mathcal{X})$ as a vector space (Lemma 1),
3) $g \in H^{\tilde{\tau}}(\mathcal{X})$, for some $\tilde{\tau} = n + r$, with $n \in \mathbb{N}$, $n \geq 2$, and $0 \leq r < 1$, where $\tilde{\tau}$ is the smoothness of $g$, and
4) the estimated smoothness of $\bar{g}$ represented as $\tau$ satisfies $\tau = n' + r'$ with $n' \in \mathbb{N}$, $n' > 1$, and $0 \leq r' < 1$, then there exists a constant $C$, which is independent of $g$, $\theta$, and $n$, such that for any $\gamma \leq \tilde{\tau}$,

$$\|g(\mathbf{x}) - \mu(\mathbf{x})\|_{H^\gamma(\mathcal{X})} \leq C h_{\mathbf{X}_{trn}^i, \mathcal{X}}^{\min\{\tilde{\tau}, \tau\} - \gamma}$$

$$\times \rho_{\mathbf{X}_{trn}^i, \mathcal{X}}^{\max\{\tau - \tilde{\tau}, 0\}} \left( \|\bar{g}\|_{H^{\tilde{\tau}}(\mathcal{X})} + \|m_\theta\|_{H^{\tilde{\tau}}(\mathcal{X})} \right)$$

The term $h_{\mathbf{X}_{trn}^i, \mathcal{X}}^{\min\{\tilde{\tau}, \tau\} - \gamma}$ decreases as more measurements are added since $h_{\mathbf{X}_{trn}^i, \mathcal{X}}$ decreases with decreasing fill distance and $\gamma$ can be selected small enough to make the exponent positive. The term $\rho_{\mathbf{X}_{trn}^i, \mathcal{X}}^{\max\{\tau - \tilde{\tau}, 0\}}$ is either constant or decreasing since the estimated smoothness is bounded and fixed for a fixed set of hyperparameters $\theta$. The terms inside the parenthesis do not depend on the estimated smoothness of the GP model. Therefore, relocation of measurement locations inside a fixed compact domain $\mathcal{X}$ does not affect boundedness and monotonic decrease of the fill distance, and as a result, does not affect convergence of the error as more measurements are added to the GP model.

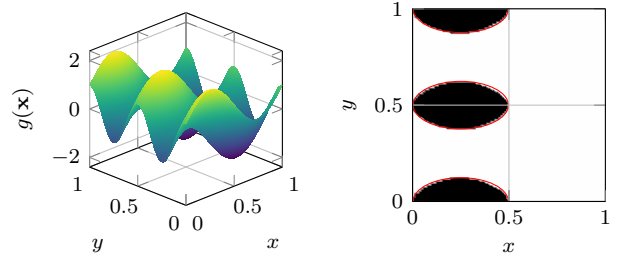

(a) 3D Surface Plot of the true scalar field $g$ used in the numerical experiment.

(b) Top view of the scalar field highlighting the high-intensity regions in black color.

Fig. 1: 3D surface plot of the true scalar field and a map of the high-intensity regions where $g(\mathbf{x}) \geq 1$.

## V. NUMERICAL EXPERIMENTS

The developed algorithm is tested using simulation in MATLAB. The scalar field is created using a continuous function $g(\mathbf{x}) = \sin(2\pi x) + \cos(4\pi y)$, where $x$ and $y$ coordinates range between 0 and 1. Therefore, the scalar estimates are bounded between [-2, +2]. This function creates multiple high-intensity regions inside the field above a threshold value of $g_{thresh} = 1$. The contours of the scalar field and high-intensity regions are shown in Figure 1a and 1b respectively.

Selection of hyperparameters in an unknown scalar field is challenging. An online routine that uses a fraction of the measurements, set aside for periodic hyperparameter optimization [13, see Chapter 5], can be run in parallel with the developed GP training method [25]. The development and implementation of such a routine is out of the scope of this paper. In this paper, we assume that using bounds on the gradient of the unknown field (estimated by a domain expert), the hyperparameters can be optimized a priori by experimenting with a set of test scalar fields with similarly bounded gradients.

A team of $N = 4$ UAVs is deployed at random starting locations inside the field with 100 randomly distributed measurement locations $\mathbf{X}_{trn}^i$ and a fixed set of 10,000 testing locations $\mathbf{X}_{test}$. Each agent takes a single measurement and calculates the mean vector and covariance matrix using (6) and (7). The hyperparameters used for this simulation are set to $\alpha = 1$ and $\beta = 0.1$. A binary image of the predicted scalar field is created using (8). A circular Hough transform is then performed on this binary image using the routine "imfindcircles" in MATLAB. This function requires a data matrix, a radius range, and a sensitivity parameter as inputs. In this simulation, the radius range of 0.05 to 0.15 and the sensitivity parameter = 1 is used. The function fits multiple circles on the high-intensity map and outputs the location coordinates and radius of the fitted circles. All future measurement locations are evaluated for presence inside high-intensity circles and relocated outside the circles.

Figure 2 shows the learned field model and high-intensity map by Agent 1 at the end of the experiment, i.e., the completion of $n_{trn}$ measurements. Figure 3 shows the de-

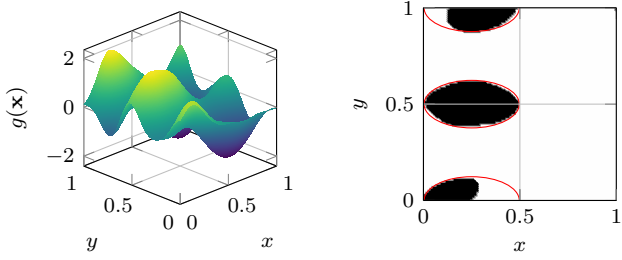

(a) 3D Surface Plot of the scalar field as estimated by Agent 1.

(b) Binary high-intensity region map as estimated by Agent 1.

Fig. 2: 3D surface plot of the scalar field as well as the high-intensity map learned by Agent 1 at the end of the experiment

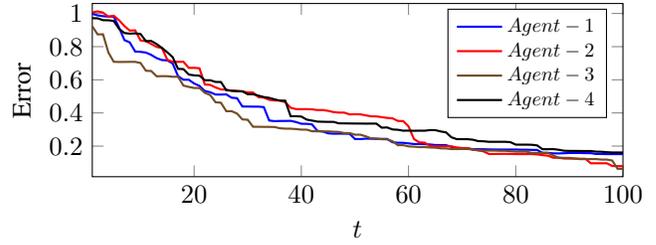

Fig. 5: The error trajectory between the true and predicted scalar field plotted against measurement iterations.

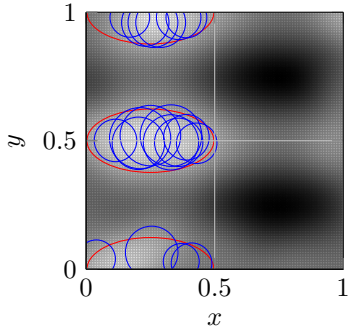

Fig. 3: This figure shows the HT fitted circles to overlap the regions of high-intensity

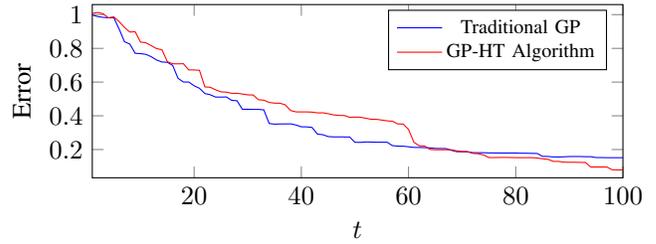

Fig. 6: Error Comparison of conventional GP and the developed GP-HT algorithm

tected circles by HT algorithm to approximate the high-intensity regions. To visualize the relocation of measurement locations, the initial and final locations are shown in Figures 4a and 4b respectively. Similar performance trends are also observed for all other agents. All agents successfully mapped the high-intensity regions with high accuracy as depicted by the error trajectory in Figure 5.

A comparative study is performed against traditional GP with no mechanism to minimize measurements from high-intensity regions. The same set of initial measurement locations is used for both methods to isolate the difference in performance due to the relocation of points inside the high-

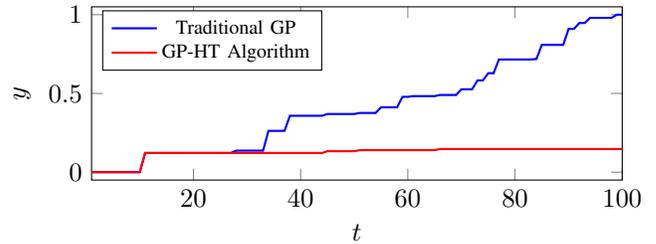

Fig. 7: The total intensity collected by Agent 1 in high-intensity region using the GP and GP-HT algorithm.

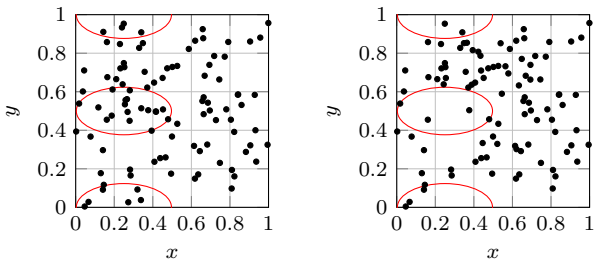

(a) Initially planned measurement locations for Agent 1.

(b) Executed measurement locations for Agent 1.

Fig. 4: Initial and final measurement locations plotted as black dots on the top view of the field.

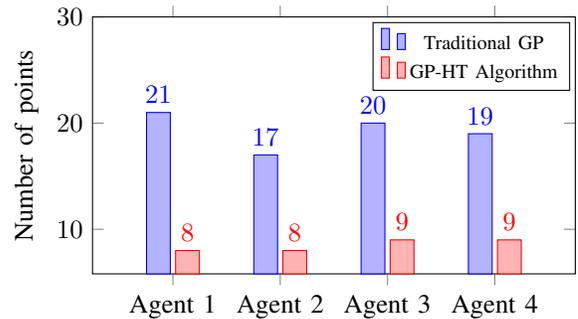

Fig. 8: Comparison of the total number of initial and final measurement locations inside all high-intensity regions.

intensity regions. Figure 6 and Figure 7, compare the overall error between the predicted field and the actual field and the total intensity measured inside the high-intensity regions respectively. The total intensity measured is the summation of field intensity measured inside the high-intensity regions. Figure 8 compares the total number of executed measurement locations inside the true high-intensity regions for both methods at the culmination of each experiment. Analysis of these plots reveals that the GP-HT algorithm results in a significant reduction of the total intensity collected inside the high-intensity regions while minimally affecting the accuracy of the learned scalar field.

## VI. Discussion

Integration of GP with the HT algorithm enables the mapping of unknown scalar fields with multiple regions of high intensity. HT-based relocation enhances agent safety, balancing exploration and safety. As shown in Section Section IV, this relocation technique does not compromise GP model efficacy.

For perfect safety, i.e., zero exposure to high field intensities, measurement locations inside the high-intensity regions should be reduced to zero. Without prior information, the GP model starts with a zero estimate, likely placing some initial measurements in high-intensity regions. Early estimates can lead to some measurements remaining in high-intensity regions. We hypothesize that in cases where partial prior knowledge of the scalar field is available, initializing the GP with an appropriate non-zero mean function could significantly improve the accuracy of high-intensity region maps, potentially preventing agents from encountering hazardous areas during the initial measurements.

## VII. Conclusion

This paper presents an approach for evading high-intensity regions inside an unknown scalar field while accurately mapping the field via the integration of a GP with an online high-intensity avoidance algorithm using the HT. The research focuses on learning of the contours of an unknown scalar field while keeping the agents safe from high-intensity regions. The simulation results indicate that a careful relocation strategy for the measurement locations inside the high-intensity regions can enhance the safety of the agents without effecting on the learning performance of the GP model. Future work will concentrate on articulating regret bounds for Gaussian Process (GP) bandit optimization, with a focus on experimental designs that enhance agent safety while expediting the learning of the GP model inside high-intensity regions.

## References

[1] P. Sabella, "A rendering algorithm for visualizing 3D scalar fields," *Comp. Graph.*, vol. 22, no. 4, p. 5158, Jun. 1988.

[2] Z. Lin, H. H. T. Liu, and M. Wotton, "Kalman filter-based large-scale wildfire monitoring with a system of UAVs," *IEEE Trans. on Indust. Electronics*, vol. 66, no. 1, pp. 606–615, Jan. 2019.

[3] K. Matsuda, K. Nozaki, R. Schregle, K. Takase, K. Taniguchi, and N. Yoshizawa, "New assessment method for solar radiation effects on indoor thermal comfort based on scalar irradiance - using volume photon mapping," *Build. Environ.*, vol. 243, p. 110662, 2023.

[4] M. T. Nguyen and K. A. Teague, "Random sampling in collaborative and distributed mobile sensor networks utilizing compressive sensing for scalar field mapping," in *10th Syst. Syst. Engg. Conf. (SoSE)*, 2015, pp. 1–6.

[5] B. Zhang and G. S. Sukhatme, "Adaptive sampling for estimating a scalar field using a robotic boat and a sensor network," in *Proc. IEEE Int. Conf. Robot. Autom.*, 2007, pp. 3673–3680.

[6] F. Dong, K. You, and X. Li, "Coordinate-free control for the isoline tracking of an unknown scalar field," *Syst. Control Lett.*, vol. 173, p. 105483, 2023.

[7] A. Dossing, E. L. S. Silva, G. Martelet, T. M. Rasmussen, E. Gloaguen, J. T. Petersen, and J. Linde, "A high-speed, light-weight scalar field magnetometer bird for km scale uav magnetic surveying: On sensor choice, bird design, and quality of output data," *Remot. Sensing*, vol. 13, no. 4, 2021.

[8] C. Wang, F. Ma, J. Yan, D. De, and S. K. Das, "Efficient aerial data collection with uav in large-scale wireless sensor networks," *Int. J. Distrib. Sens. Netw.*, vol. 11, no. 11, p. 286080, 2015.

[9] R. A. Razak, S. Sukumar, and H. Chung, "Estimating scalar fields with mobile sensor networks," in *6th Ind. Control Conf.*, 2019, pp. 63–68.

[10] H. M. La, W. Sheng, and J. Chen, "Cooperative and active sensing in mobile sensor networks for scalar field mapping," *IEEE Trans. Syst. Man Cybern. Syst.*, vol. 45, no. 1, pp. 1–12, 2015.

[11] R. Cui, Y. Li, and W. Yan, "Mutual information-based multi-AUV path planning for scalar field sampling using multidimensional RRT," *IEEE Trans. Syst., Man and Cybern. Syst.*, pp. 993–1004, 2016.

[12] F. Lian, A. Chakrabortty, and A. Duel-Hallen, "Game-theoretic multi-agent control and network cost allocation under communication constraints," *IEEE J. Sel. Areas Commun*, pp. 330–340, 2017.

[13] C. E. Rasmussen and C. K. I. Williams, *Gaussian processes for machine learning*. Cambridge, MA: MIT Press, 2006.

[14] T. X. Lin, S. Al-Abri, S. Coogan, and F. Zhang, "A distributed scalar field mapping strategy for mobile robots," in *Intl. Conf. on Intel. Rob. Syst.*, 2020, pp. 11 581–11 586.

[15] G. Antonelli and S. Chiaverini, "Kinematic control of platoons of autonomous vehicles," *IEEE Trans. Robot.*, vol. 22, no. 6, pp. 1285–1292, 2006.

[16] B. Bayat, N. Crasta, A. Crespi, A. M. Pascoal, and A. Ijspeert, "Environmental monitoring using autonomous vehicles: a survey of recent searching techniques," *Curr. Opin. Biotechnol.*, vol. 45, pp. 76–84, 2017, energy biotechnology Environmental biotechnology.

[17] D. Casbeer, R. Beard, T. McLain, S.-M. Li, and R. Mehra, "Forest fire monitoring with multiple small uavs," in *Proc. of Amer. Control Conf.*, 2005, pp. 3530–3535 vol. 5.

[18] J. Aleotti, G. Micconi, S. Caselli, G. Benassi, N. Zambelli, M. Bettelli, and A. Zappettini, "Detection of nuclear sources by uav teleoperation using a visuo-haptic augmented reality interface," *Sensors*, vol. 17, no. 10, 2017.

[19] S. Okuyama, T. Torii, A. Suzuki, M. Shibuya, and N. Miyazaki, "A remote radiation monitoring system using an autonomous unmanned helicopter for nuclear emergencies," *J. Nucl. Sci. Technol.*, vol. 45, no. sup5, pp. 414–416, 2008.

[20] B. Sangiovanni, G. P. Incremona, M. Piastra, and A. Ferrara, "Self-configuring robot path planning with obstacle avoidance via deep reinforcement learning," *IEEE Control Syst. Lett.*, vol. 5, no. 2, pp. 397–402, 2021.

[21] M. T. Nguyen, H. M. La, and K. A. Teague, "Compressive and collaborative mobile sensing for scalar field mapping in robotic networks," in *53rd Annual Allert. Conf. on Comm., Cont., and Comput.*, 2015, pp. 873–880.

[22] R. O. Duda and P. E. Hart, "Use of the hough transformation to detect lines and curves in pictures," *Communications*, vol. 15, no. 1, pp. 11–15, Jan. 1972.

[23] A. L. Teckentrup, "Convergence of Gaussian process regression with estimated hyper-parameters and applications in bayesian inverse problems," *SIAM/ASA J. Uncertainty Quantif.*, vol. 8, no. 4, pp. 1310–1337, 2020.

[24] H. Wendland, *Scattered Data Approximation*, 1st ed. Cambridge, MA: Cambridge University Press, 2004.

[25] J. Wu, X.-Y. Chen, H. Zhang, L.-D. Xiong, H. Lei, and S.-H. Deng, "Hyperparameter optimization for machine learning models based on bayesian optimizationb," *Journal of Electronic Science and Technology*, vol. 17, no. 1, pp. 26–40, 2019.